\documentclass[12pt]{article}
\usepackage{mathtools}
\usepackage{hyperref}
\usepackage{amsmath}

\begin{document}

\title{The magical ``Born Rule" \& quantum ``measurement": Implications for Physics}
\author{Johan Hansson\footnote{\href{mailto:c.johan.hansson@ltu.se}{c.johan.hansson@ltu.se}} \\
 \textit{Division of Physics} \\ \textit{Lule\aa  University of Technology}
 \\ \textit{SE-971 87 Lule\aa, Sweden}}

\date{}

\maketitle

\begin{abstract}
I. The arena of quantum mechanics and quantum field theory is the abstract, unobserved and unobservable, $M$-dimensional formal Hilbert space $\neq$ spacetime.
II. The arena of observations and, more generally, of \textit{all} events (\textit{i.e. everything}) in the real physical world, is the classical 4-dimensional physical spacetime.
III. The ``Born Rule" is the random process ``magically" transforming I. into II.
Wavefunctions are superposed and entangled only in the abstract space I., \textit{never} in spacetime II. Attempted formulations of quantum theory directly in real physical spacetime actually constitute examples of ``locally real" theories, as defined by Clauser \& Horne, and are therefore \textit{already empirically refuted} by the numerous tests of Bell's theorem in real, controlled experiments in laboratories here on Earth.
\textit{Observed} quantum entities, \textit{i.e.} events, are \textit{never} superposed or entangled as they: 1) Exclusively ``live" (manifest) in real physical spacetime, 2) Are not described by entangled wavefunctions after ``measurement", effectuated by III.
When separated and treated correctly in this way, a number of fundamental problems and ``paradoxes" of quantum theory vs. relativity (\textit{i.e.} spacetime) simply vanish, such as the black hole information paradox, infinite zero-point energy of quantum field theory and quantization of general relativity.
\end{abstract}

\section{Introduction: Quantum to Classical}
I. Quantum theory lives in abstract Hilbert space: $H^M$, which, more often than not, is infinite-dimensional $H^{\infty}$ \cite{vonNeumann}. Complex quantum wavefunctions, $\psi$ (= vectors in $H^M$), for $N$ discrete quantum entities are defined in \textit{configuration} space of $3N$ dimensions (if their spins are zero), $\psi = \psi (q_1, ..., q_{3N})$. The quantum state, \textit{i.e.} the value of $\psi$, is determined by simultaneously giving all numerical values of the $3N$ variables $(q_1, ..., q_{3N})$. The time-dependence of $\psi$ is only implicit, as $t$ neither is an operator nor a variable in configuration space, but merely a parameter, where $\psi_t = \exp^{-itH/\hbar} \psi_0$, and this ``dynamics" occurs in abstract, complex Hilbert space, \textit{not} in spacetime. There is no spacetime description of $\psi$, so relativistic causality is not definable. The quantum states are normalized, $\int |\psi (q_1, ..., q_{3N})|^2 dq_1 ... dq_{3N} = 1$ (\textit{i.e.} $\psi$ lie at a point on the surface of the unit sphere in Hilbert space).

II. Classical physics lives in four-dimensional physical spacetime: the Lorentzian manifold $L^4$.

III. On ``measurement" the ``Born Rule" \cite{Born} is \textit{postulated} to irreversibly, instantaneously and randomly map $H^{M}$ into specific points (= \textit{events, i.e.} particular outcomes) in $L^4$, with calculable probabilities. The joint probability density of \textit{finding} $N$ ``particles" in $N$ \textit{detectors} in \textit{real} space at time $t$ (\textit{i.e.} in $L^4$) upon \textit{``measurement"} is calculated by $|\psi (\bar{r}_1, ..., \bar{r}_N)|^2$. Observe that there is \textit{no} real spatial dependence for $\psi$ until this ``measurement". The ``fundamental" statistical character of ``quantum theory" is actually only introduced here in III., due to ``Born". Everything we ever measure (\textit{e.g.} using laboratory ``detectors"), perceive or experience occurs in \textit{spacetime}, but it is only the Eigenvalues that can be observed in $L^4$, the quantum Eigenfunctions still reside in $H^{M}$ (they constitute bases there) even after ``Born". The ``Born Rule" is ``magical" in the sense that there is no physical dynamics underlying it \textit{and} that it transforms \textit{unobservable} $H^M$ into \textit{observable} events in $L^4$. On ``measurement", the wavefunction $\psi$, with complex quantum amplitudes $c_n = \langle n | \psi \rangle$ in Eigenbase $| n \rangle$, randomly ``collapses" (``jumps") to \textit{one} of the Eigenfunctions $| n \rangle$ of the, for observables, allowed Hermitian operators $\hat{O}_n$ (assuring real Eigenvalues $o_n$)

\begin{equation}
\hat{O}_{n} \psi = \hat{O}_{n} \sum_n |n \rangle \langle n | \psi \rangle \xrightarrow{\text{``meas"}} o_{k} c_{k} | k \rangle.
\end{equation}
Each individual ``jump", $\psi \rightarrow | k \rangle$, resulting in the observed Eigenvalue, $o_k$, is \textit{a priori} postulated to occur randomly, but with statistical probability (over \textit{many} identical measurements) given by $|c_k |^2 =  |\langle k | \psi \rangle |^2$. It is not the operators themselves that are observed, they only operate in Hilbert space \textit{not} in spacetime. It is the Eigenvalues that are observed, and then only indirectly as a result of ``measurement", Eq. (1), transforming I. into real events in II. as a result of III.

However, we only infer I. and III. \textit{indirectly} through observations, experiments and experiences in II. - the only world we have direct access to. \textit{Observed} observables (= events) live in $L^4$. Quantum entities solely live in $H^{M}$ and \textit{have no classical properties}. The ``measurement" transforms the infinitely many abstract quantum potentialities in $H^M$ into perfectly mundane actual occurrences in $L^4$. Consequently, this shows that ``decoherence" \cite{Zurek} \textit{cannot} be a solution to the ``measurement problem" \cite{Bell3}, \cite{Adler}, \cite{BellAgainst}, \cite{Wigner2}, \cite{Wigner}, in quantum mechanics: i) It does not realize \textit{any} objective outcome (unlike ``Born"), ii) And if it did, it would mean that our classical world would be \textit{manifestly non-}local (as decoherence is based purely on the wholly deterministic non-local ``dynamics" of I.) In pure quantum theory (without ``Born") there cannot \textit{be} any mixed states, as \textit{probability presumes} prior measurement.

Hence we see that statistically correlated \textit{observed} ``quantum" non-locality in spacetime in practice actually \textit{only} arises through the magical ``Born Rule". However, it still poses serious problems, as the \textit{results} of measurement (\textit{after} ``Born") are \textit{objective classical} events in $L^4$ (\textit{e.g.} data printouts on paper), \textit{but} ``simultaneous" is \textit{not} relativistically invariant: In a canonical entangled-pair experiment, with correlated observables A and B at either end, for an observer moving relative to the lab (with any non-zero velocity $v$, however small, $\mid v \mid = \epsilon > 0$) A is prior to B if $v > 0$ (A is the ``cause" of B and the ``Born collapse" is \textit{not} instantaneous), but B is prior to A if $v < 0$ (B is the ``cause" of A and the ``Born collapse" is \textit{not} instantaneous), if A is simultaneous with B in the lab-frame ($v=0$) \cite{HanssonPhysEssays}. The problem is that the ``Born Rule" is formulated in an \textit{absolute} frame, the one where $v=0$.

``Reality" occurs \textit{only} in the spacetime of events - where the actual events are the fundamental, relativistically invariant and irreducible building blocks of objective reality \cite{Wheeler}. It is only here, in II., that all experimental results, and everything else we ever perceive, actually occur. That is why Bohr was fond of saying: ``There is no quantum world" \cite{Bohr} only an abstract quantum \textit{algorithm}, I. together with III., allowing us to relate experiences in the real world II. - the only one. There is no ``quantum" reality, there are no ``quantum" events, only a \textit{classical reality} and \textit{classical events}. Every time a ``quantum" probability is calculated, it is really a result of ``Born" III., \textit{not} of pure (unmeasured) quantum theory I. Likewise, there are no ``quantum" particle reactions in \textit{spacetime}, only observed consequences in $L^4$. So, only II. is really real, I. and III. merely an abstract, and unobservable machinery, very much like a black box we cannot peer into, but with observable inputs and outputs. In a very real sense, the ``quantum world" is operationally built up of events in our real world, not the other way around. Reality does \textit{not} occur in Hilbert space $H^M$. This also means that there are no fundamental quantum entities in \textit{spacetime}, only in (unobservable) Hilbert space.

The \textit{interaction} in quantum theory I. is \textit{non-local} in non-relativistic quantum mechanics (\textit{e.g.} Schr\"{o}dinger) and \textit{local} in relativistic quantum theory (Dirac/quantum field theory), but the wavefunction is \textit{non-local} in \textit{both} (but only implicitly, in abstract Hilbert space, not in spacetime).

Wavefunctions are entangled \cite{Schrodinger} only in Hilbert space, \textit{not} in real physical spacetime.

The dynamical real spacetime itself (II.) is local. The entanglement superposition (in $H^M$) is \textit{broken} by the measurement, I. $\xrightarrow{\text{III.}}$ II., hence there is \textit{never} any non-causal \textit{entanglement} in real spacetime $L^4$. Only classical particles and fields are defined directly in $L^4$.

III. is \textit{non-local} \cite{EPR} in real \textit{spacetime}. It correlates space-like separated events in our \textit{real world}. But particles ``manifest" as events only as a result of ``measurement" (through the ``Born rule") - it is therefore fundamentally \textit{wrong} to assume that they separate and travel, moving apart in real physical space, to the detectors from the source while unobserved. (In Hilbert space they \textit{neither} separate \textit{nor} are ``far apart" as they are described by the \textit{global} wavefunction in $H^M$.) The Bohm version \cite{BEPR} of the EPR-gedankenexperiment \cite{EPR} \textit{disregards} the actual ``measurement", as the \textit{spatial} part ($\psi_{space}$) of the \textit{total} wavefunction is omitted/implicit. As we know today, it \textit{cannot} be factored, $\psi_{tot} = \psi_{tot} (\bar{q}_1, \bar{q}_2, \bar{s}) \neq \psi_{space} \psi_{spin}$, as $\psi_{tot}$ is \textit{global} and depends on both quantum entities in an entangled, \textit{not} factorizable, way, $\psi_{space} = \psi_{space}(\bar{q}_1, \bar{q}_2)$, $\psi_{spin} = (|\uparrow\rangle_{\bar{q}_1} |\downarrow\rangle _{\bar{q}_2} - |\downarrow\rangle_{\bar{q}_1} |\uparrow\rangle _{\bar{q}_2})/\sqrt{2}$. Only quantum entities that do \textit{not} interact \textit{and have never} interacted may be factorized. The ``measurement" (``Born Rule") collapses \textit{both} space- \cite{EPR} and spin-parts \cite{BEPR} of $\psi_{tot}$ (which due to the spin-1/2 degree of freedom in this case lives in $H^{12})$ \textit{at once}.

Unobserved quantum entities are always (merely abstract) ``waves" \textit{in $H^M$}, observed quantum entities are always ``particles" manifested as events \textit{in $L^4$} - \textit{there is never any ``particle-wave-duality" in either space}. Specifically, there is never any causal ``quantum-wave" propagation in spacetime. This means that classical physics, II., can \textit{never} be the limit $\hbar \rightarrow 0$ of ``pure" quantum theory, I.

The quantum description of a system of $N$ entities (for $N > 1$) cannot be embedded in real spacetime \cite{ClauserConfig} - actually, the very formulation of quantum mechanics \textit{precludes} its embedding in spacetime for $N > 1$. For example, the quark- and gluon-fields ($M = \infty$) interacting ``in" a proton never objectively exist as particles in spacetime - only when a ``measurement" is made, for example using deep inelastic scattering, the results of an ``electron-quark collision", mediated by a photon- or $Z^0$-\textit{field} (\textit{not} particle), in Hilbert space ($H^{\infty}$) through ``Born" becomes translated into some experimental signal in real 4D spacetime ($L^4$).

Abstract configuration space $(q_1, ..., q_{3N})$ and physical space $(x, y, z)$ can coincide \textit{only} if there is only \textit{one} (spinless) quantum entity, actually \textit{measured} at $(x, y, z)$, in the entire universe (this unfortunately \textit{precludes} any interactions, experiments and observers), otherwise they are distinct - and actually the origin of most confusion.

``Born", III., is just a random sampling, upon \textit{``measurement"}, of the abstract, globally ever present and completely \textit{deterministic} Hilbert space - meaning that the (unobservable) ``quantum world" is completely deterministic - determined by all the \textit{unobserved} variables in configuration space, while the real world is local and uncertain in part due to our ignorance of the global/non-local ``hidden variables" of configuration space.

We thus see that even \textit{orthodox} quantum mechanics already, in a sense, \textit{has} ``hidden variables" in fact always present in configuration space, which globally keeps track both of what has happened, and also of everything that can ever happen - potentially including even the ``decisions" of observers.

Even for two \textit{free} quantum entities, that have \textit{ever} interacted in the past, measurement on one affects measurement on the other. For example, in energy Eigenbase
\begin{equation}
\psi = A \psi_a (\bar{q}_1) \psi_b (\bar{q}_2) + B \psi_c (\bar{q}_1) \psi_d (\bar{q}_2)
\end{equation}
an energy Eigenvalue measurement on particle one depends on the energy measurement on the other, regardless of their separation in $L^4$ (this being just a special example of \textit{entanglement} of two presently \textit{non-interacting} quantum entities), where $|A|^2 + |B|^2 = 1$. This entanglement \textit{persists} indefinitely until ``measurement" $\Rightarrow$ ``Born" $\Rightarrow$ ``collapse of the wavefunction" $\Rightarrow$ probability ensembles \textit{in spacetime}, $L^4$.

\section{Explicit Collapse vs. ``Magical" Collapse}
In explicit collapse models, the standard linear quantum evolution is complemented by (\textit{ad hoc}) nonlinear terms which become important for macroscopic entities, inducing (stochastic) dynamical collapse, \cite{Pearle}, \cite{GRW}, \cite{Penrose}, \cite{Diosi}. One version uses the already present nonlinearities in non-abelian quantum field theory \cite{HanssonCollapse}.

The \textit{dis}entanglement in ``orthodox" quantum mechanics is only produced by active ``measurement" (observation): ``Born", and ``measurement" on one side, $\bar{x}_1$, results in instantaneous ``measurement" on the other side, $\bar{x}_2$ - regardless of spatial distance $|\bar{x}_1 - \bar{x}_2|$. This means that ``Born" must break relativistic invariance as: i) ``Simultaneous events" is not a relativistically invariant concept. ii) Any, arbitrarily small, non-zero distance $\epsilon = |\bar{x}_1 - \bar{x}_2| > 0$ is automatically spacelike (relativistically non-causal) if $t_1 = t_2$, \textit{i.e.} if the ``collapse" is instantaneous.

In dynamical collapse models, energy is not strictly conserved. There exists no continuous Noether symmetry in time, as the dynamical collapse is irreversible. This is side-stepped in ``Born" collapse as it is \textit{non}-dynamical - ``magical".

Even if the actual ``measurement" is assumed to take a finite time, we still obtain a causal paradox as the ``measurement" at the other end is \textit{not} connected to the first by a Lorentz transformation if the opposite ends are spacelike-separated in spacetime \cite{HanssonPhysEssays}.

\section{Linearity vs. Non-linearity \& \\ Locality vs. Non-locality}
\subsection{Quantum Theory}
Quantum theory for $N$, even non-interacting, spinless quantum entities lives in $H^{3N}$, an abstract, complex, \textit{linear} (vector) space. The evolution in $H^{3N}$ is continuous, linear \cite{Berry}, reversible, non-local (but merely abstractly/implicitly so) and deterministic (describable by differential equations). The wavefunction is not defined until/unless all points in configuration space $(q_1, ..., q_{3N})$ are used as input. For $N > 1$, quantum theory cannot be embedded in real physical spacetime $L^4$ \cite{ClauserConfig}. The spacetime description is \textit{only} appropriate for our \textit{detectors and observations} in $L^4$ - not for the abstract theory supposedly ``underlying it all" in $H^{M}$. No quantum fields ever ``permeate" spacetime.
\subsection{Classical Physics}
Events define, and also constitute, dynamical classical spacetime, $L^4$.
The dynamics is continuous, nonlinear, reversible, local, causal and deterministic (describable by generally relativistic covariant differential equations). This \textit{nonlinear} dynamics evidently \textit{cannot} result from ``pure" linear quantum theory alone.
\subsection{The magical ``Born Rule"}
The ``Born Rule" is \textit{discontinuous}, nonlinear, \textit{irreversible} (entropy increasing \cite{vonNeumann}), non-local (explicitly - assumed to be instantaneous in \textit{spacetime}), non-causal and \textit{postulated} to be intrinsically/fundamentally \textit{random}/probabilistic (\textit{e.g.} giving no possibility of superluminal \textit{signalling} despite the, now physical, non-locality in spacetime). It is \textit{not} describable by differential equations, or in any other dynamical way, instead being ``magical". Observe that ``Born" kills all superpositions (including entanglements) as the end result is a classical \textit{probability} $\propto | \psi|^2 $, not longer any interfering amplitudes/wave functions. This also means that there can be no superpositions in spacetime (or \textit{of} spacetimes), as probabilities do not interfere, only add, forbidding any ``quantum spacetime". It maps $H^{M} \xrightarrow{\text{``Born"}}$ into specific \textit{outcomes} (= events) in spacetime, $L^4$. Observe that the \textit{Eigenvalues} are the physical (and random) ``observables" in $L^4$, \textit{never} the Eigenfunctions themselves (they perpetually live in abstract, complex Hilbert space). Expectation values, $\langle \hat{O} \rangle = \langle \psi | \hat{O} | \psi \rangle$, are statistical averages of many measured Eigenvalues in identically ``prepared" systems, $| \psi \rangle$, and are predictable in a statistical sense only.

As Bell showed, \textit{all} measurement results can ultimately be boiled down to \textit{position} results \cite{BellPosition} which, together with time, \textit{are} the events in spacetime, $L^4$.

A hypothetical free (non-interacting and spinless) single particle can be represented in physical spacetime \textit{only} when $(q_1, q_2, q_3) = (x, y, z)$ = the location of the \textit{detector in} $L^4$ at real \textit{``measurement"} of the particle, and then by an infinite wave-train with equal probability (= 0) to be \textit{anywhere} (at ``measurement"). If instead regarded as semi-localized wave-packets (infinite superposed sum of different wave-trains) they will: i) disperse, ii) not have a unique energy or propagation speed, meaning that there would be no reason they should arrive at a detector at a calculable time. Hence, even single quantum ``particles" \textit{cannot} travel \textit{through} spacetime as microscopic ``bullets". The momentum ``conservation" always assumed, \textit{e.g} to ensure spatial correlation of entangled pairs, actually occurs in abstract Hilbert space, \textit{not} physical spacetime. Neither particle in the ``pair" exists \textit{anywhere} in spacetime until/unless ``measured". Quantum ``particles" have no trajectories in spacetime, and if $N > 1$ the evolution \textit{cannot} be embedded in spacetime anyway.


From quantum theory alone, there is thus \textit{no reason} that detection of both ``particles" of an entangled pair should be detected simultaneously, equidistant from the source. This can, at best, hold only \textit{in the mean} as: i) each \textit{individual} measurement event is random (\textit{postulated} so by ``Born"), ii) the probability of ``Born" are weighted statistical means of very many individual (random) measured events.


The \textit{locality} assumption only applies to real physical spacetime, \textit{not} to abstract Hilbert space where obviously everything is non-locally interconnected through the global configuration - \textit{i.e.}, ``unmeasured" quantum theory does not respect Lorentz invariance - but this is irrelevant as Lorentz invariance is only observed in \textit{spacetime} and $H^{M}$ itself is \textit{unobservable} in principle.

\section{Quantum Space $\neq$ Real Spacetime: \\ Some Physical Consequences}
The non-locality in Hilbert space is an abstract ``unphysical", ever-present, global non-locality. But it becomes a non-locality in real physical spacetime through ``Born's Rule". The non-locality of measurement is evident already in the 1-particle case, as pointed out already very early on by Einstein \cite{EinsteinBubble}, but it becomes \textit{experimentally testable} in $N$-particle entangled states. Originally, tests had $N = 2$, \cite{Bell2}, \cite{BellTests}, and \textit{all} ``locally real" \cite{ClauserHorne} models formulated in real spacetime, $L^4$, are soundly \textit{falsified} by these tests \cite{ClauserConfig}, \textit{including} quantum mechanics and quantum field theory \textit{formulated in real spacetime}. Hence, a truly relativistically invariant formulation of quantum theory in \textit{spacetime}, which \textit{includes} ``measurement", could never be compatible with the non-locality of nature already observed in these tests, as correlations in real outputs of real experiments in our real world II.
\subsection{Consequence 1: No ``Quantized" General \\ Relativity}
Apart from having completely different mathematical structures, quantum theory and general relativity ``live" in completely different spaces, which means that ``Quantum General Relativity" and ``Quantum Spacetime" are meaningless concepts \cite{GravNonQuantum}. Quantum theory lives in the abstract mathematical linear vector space $H^{M}$ with perfectly deterministic, and linear evolution. General relativity lives in, and actually constitutes, real physical 4-dimensional spacetime $L^4$ with non-linear causal evolution of chains of ``events" = the actual ``happenings" that constitute the fundamental, irreversible, invariant ``constituents" of spacetime, which, when warped by \textit{classical} energy-momentum in \textit{spacetime}, $T^{\mu} _{\nu}$, results in \textit{classical} gravitation in $L^4$ through Einstein's equations $G^{\mu} _{\nu} = \kappa T^{\mu} _{\nu}$.
\subsection{Consequence 2: No ``Zero-Point Energy" or \\ Cosmological Constant Problem}
Virtual ``particles" exclusively live in Hilbert space, \textit{not} in physical spacetime. They never manifest into $L^4$ through ``Born". That is why they are not real. The same applies for the infinite ``zero-point energy" of the quantum vacuum in quantum field theory arising from ``virtual particles". Which in turn explains why the cosmological constant, $\Lambda$, does \textit{not} go to infinity, and hence why the physical cosmos ($L^4$) has been able to expand leisurely without ripping itself apart.

As ``virtual particles" never physically manifest in spacetime they have no influence at all on the classical energy density $T^0 _0$, or pressure $T^i _i$, in spacetime, so \textit{no} effect on the expansion of the universe, given by Einstein's equations: $G_{\nu}^{\mu} = \kappa T_{\nu}^{\mu} + \Lambda g_{\nu}^{\mu} \overset{\mathrm{?}}{=} \kappa T_{\nu}^{\mu} + \kappa \langle 0 | \hat{T}_{\nu}^{\mu}(virtual) | 0 \rangle \equiv \kappa T_{\nu}^{\mu}$.\footnote{We have here assumed a $\Lambda$ that is solely due to the presently very fashionable, albeit completely \textit{hypothetical}, ``Dark Energy", \textit{i.e.} ``quantum vacuum". As $\Lambda$ in \textit{classical general relativity} is merely a free parameter we can \textit{choose} it to have any value whatsoever to comply with cosmological observations (\textit{e.g.} finite and \textit{very small}). Such a $\Lambda_{classical}$ would give a \textit{curvature} in spacetime even in the absence of $T_{\nu}^{\mu}$ (\textit{i.e.} in the \textit{classical} vacuum) but $\Lambda_{classical}$ is \textit{not} a classical vacuum energy, which by \textit{definition} is identically zero. It is a geometric curvature of empty spacetime itself.} In fact, there is \textit{no} instance where this ``vacuum energy" is actually physically needed \cite{Casimir}, \cite{Lifshitz}, \cite{Jaffe}.\footnote{While our argument is based solely on ``normal" quantum theory, another, hypothetical, and also highly abstract, theory $\neq L^4$ hints at the same conclusion \cite{Amplituhedron}.}

\subsection{Consequence 3: Quantum ``particle" reactions do \textit{not} happen in spacetime but in Hilbert space}
Only the \textit{observed} (``measured") quantum lives in spacetime, quite contrary to what one might believe when drawing innumerable, linearly superposed Feynman diagrams. A particle, in $L^4$, \textit{never} occupies two (or more) distinct positions at the same time. The quantum superpositions occur in $H^M$, where \textit{no} classical attributes ever manifest. The same goes for ``particle" interactions in particle physics, which by definition have $N > 1$, the quantum reactions happen in $H^{\infty}$, the \textit{outcomes} happen in $L^4$, and then \textit{only} as a result of ``Born". In the canonical two-slit experiment, \textit{e.g.} using a laser, each ``hit" at the detector screen is the result of a quantum (photon-field) interaction in $H^{\infty}$ manifesting as an \textit{event} in $L^4$ where only \textit{one} discrete small region of the screen randomly lights up, as if by a photon-``particle". It is only after \textit{many} hits that the superposition (in $H^{\infty}$) becomes manifested in our real world ($L^4$) as real discrete data patterns through the statistical ``Born Rule", Eq. (1).

\subsection{Consequence 4: No Black Hole ``information \\ paradox"}
As wavefunctions, for $N \geq 2$ quantum entities, are objects in Hilbert space with global entanglement through $(q_1, ..., q_{3N})$, \textit{not} in $L^4$, they are unaffected by causal horizons in spacetime, meaning that quantum entities inside the horizon are always accessible by entangled quantum entities outside the horizon - nullifying, for \textit{quantum theory}, \textit{e.g.} the classical one-way membrane of a black hole event horizon - and hence potential information is in principle always accessible across horizons. A causal probability current in spacetime is definable, and conserved, only for a single, non-interacting particle, making it physically irrelevant. For $N$ quantum entities, entangled or not, \textit{no} conserved probability current is definable in spacetime, and hence can never ``flow" causally. (And neither in Hilbert space as ``probability" requires that ``Born" already has occurred.) The abstract non-locality in Hilbert space binds arbitrarily distant quantum entities into a single global irreducible $\psi$. ``Born" then binds actual events non-locally in real spacetime, regardless of spacetime-interval separation. This resolves the quantum information paradox \cite{Qinfoparadox} for black holes, making it a non-question.
\section{Some proposed Alternatives to ``Orthodox" Quantum Mechanics}
\subsection{Everett/many worlds} Only Hilbert, no collapse \cite{Everett}. i) \textit{Linear}, cannot give the nonlinear classical world \cite{Berry}. ii) Does not give any probabilities (no ``Born"), and never even any outcomes at all, meaning that a classical world is \textit{absent} in \textit{all} parallel ``universes".

\subsection{Explicit collapse} i) \textit{Dynamical} nonlinear collapse, does not give a classical world as explicit non-locality (in principle) persists in real spacetime and energy is not strictly conserved. ii) Collapse time \textit{not relativistically invariant}, cause-effect for entangled systems ill-defined (depends on frame).

\subsection{de Broglie-Bohm} No collapse \cite{deBroglie}, \cite{Bohm}, everything is (in principle) completely deterministic. i) Does not give a classical world. ii) Positions for particles always live directly in spacetime, and are guided by an \textit{extra} equation, simultaneously the guiding ``pilot wave", $\psi$, lives in Hilbert.
The de Broglie-Bohm theory has \textit{no} need for a ``Born Rule", as the classical level is objectively real all the time, but the ``pilot wave" guiding the (now objectively real) quantum \textit{particles} is manifestly non-local and lives in unphysical Hilbert space, eternally global in configuration space, but as its predictions are designed to be exactly those of orthodox quantum mechanics it \textit{cannot} explain the nonlinearity of classical physics. Through the guiding equation (which includes $\psi$) the positions of particles in \textit{spacetime} depend on the positions of all other particles (arbitrarily far) making also the dynamics in real spacetime \textit{manifestly} non-local, \textit{i.e.} it breaks the relativistic invariance of the real world II. \textit{explicitly}. In the orthodox theory it is ``Born" III., that saves the real world II. from manifestly/deterministically breaking relativistic invariance, as ``Born" is only \textit{statistically} non-local in spacetime.
\section{Summary \& Conclusion}
``Pure" quantum theory, I., is \textit{implicitly} non-local, but the non-locality is unphysical (not observable) as it does not ``live" in spacetime but in Hilbert space.

The ``Born Rule", III., is \textit{explicitly} non-local for entangled quantum systems - it correlates spacelike separated \textit{events} in \textit{real} spacetime, as required by Bell's theorem and its empirically validated requirement of a non-local reality.

``Reality", II., occurs only in the spacetime of \textit{events} (which are the fundamental ``building-blocks" of objective reality) \textit{not} in quantum Hilbert space. Thus, ``quantum information" is a misnomer, information only manifest in \textit{spacetime} after \textit{``measurement"} (\textit{i.e.} ``Born Rule", III.) has occurred.

The fact that quantum systems, with more than one quantum entity $N>1$, \textit{cannot} be embedded in spacetime has very deep, profound and startling consequences. It means, for instance, that quarks and gluons are \textit{not} ``constituents" of (\textit{e.g.}) protons in \textit{spacetime}, only in abstract, infinite-dimensional Hilbert space \cite{Confinement} - the proton is \textit{not} a ``bag" (in spacetime) \textit{containing} quarks and gluons\footnote{This is probably the solution to the ``proton spin crisis" \cite{SpinCrisis}.}. More generally, fundamental (quantum) ``particle" interactions \textit{never} occur in spacetime: Rather, merely abstract quantum fields in $H^{\infty}$ result, through the magical ``Born Rule", in observed phenomena as objective events in real physical spacetime \textit{interpreted} as particles. Objects in our \textit{real} world $L^4$ thus do not ``consist" of fundamental quantum entities. Not even atoms or molecules ``consist" of electrons, protons and neutrons \textit{in spacetime}, rather the entangled electron-proton-neutron wavefunction in \textit{Hilbert space} can manifest as events (in $L^4$) interpreted as arising from ``atoms" and ``molecules" upon ``measurement", \textit{i.e.} upon ``Born". Even for superfluids and superconductors, macroscopic in size, the quantum properties perpetually live in $H^{M}$ alone. The \textit{observations} of superfluids/conductors are always perfectly mundane events in our normally perceived world. Also, the $10^{57}$ neutrons in a neutron star live in a configuration space of $3 \times 10^{57}$ dimensions in Hilbert (\textit{not} in spacetime) resulting, again as always through ``Born", in the observed physical properties of the neutron star in $L^4$. (Quarks in quark(-gluon) stars, if they exist at all, would live in $H^{\infty}$.) ``Schr\"{o}dinger's Cat" \cite{Schrodinger} is dead \textit{or} alive in our real world II., \textit{after} ``Born" III. (``magically") has realized the outcome from its entangled wavefunction in merely abstract Hilbert space I.

The only mystery remaining is why (and how?) the ``Born Rule" occurs at all. But then again, maybe nature really \textit{is} magical.

\end{document}